\documentclass[nofootinbib,twocolumn,pra]{revtex4-1}
\usepackage{xspace,amsmath,amsfonts,amsthm,amssymb,amsbsy,graphics,color,graphicx,bbm,soul}
\usepackage[dvipsnames]{xcolor}
\usepackage[breaklinks=true]{hyperref}

\usepackage{mathpazo}

\usepackage{epigraph}

\hypersetup{
  colorlinks   = true, 
  urlcolor     = blue, 
  linkcolor    = blue, 
  citecolor   =  red 
}

\newcommand{\bra}[1]{\left\langle{#1}\right|}
\newcommand{\ket}[1]{\left|{#1}\right\rangle}
\newcommand{\op}[2]{\ket{#1}\!\!\bra{#2}}

\newtheorem{thm}{Theorem}

\begin{document}
\title{Classical correlation alone supplies the anomaly to weak values}
\author{Christopher Ferrie}

\affiliation{
Center for Quantum Information and Control,
University of New Mexico,
Albuquerque, New Mexico, 87131-0001, USA}

\author{Joshua Combes}
\affiliation{Institute for Quantum Computing, University of Waterloo, Waterloo, Ontario N2L 3G1, Canada}
\affiliation{Perimeter Institute for Theoretical Physics, 31 Caroline St. N, Waterloo, Ontario N2L 2Y5, Canada}

\begin{abstract}
The question of what is genuinely quantum about weak values is only ever going to elicit strongly subjective opinions---it is not a scientific question.  Good questions, when comparing theories, are operational---they deal with the unquestionable outcomes of experiment.  We give the anomalous shift of weak values an objective meaning through a generalization to an operational definition of anomalous post-selected averages.  We show the presence of these averages necessitate correlations in every model giving rise to them---quantum or classical.  Characterizing such correlations shows that they are ubiquitous.  We present the simplest classical example without the need of disturbance realizing these generalized anomalous weak values.
\end{abstract}

 \date{\today}

\maketitle

\epigraph{The combination of some data and an aching desire for an answer does not ensure that a reasonable answer can be extracted from a given body of data.}{John W. Tukey \cite{tukey}}

\section{Introduction}
After many years, the fringe topic of \emph{weak values} \cite{Aharonov1988How} has entered the spotlight of mainstream physics \cite{wv_review}.  Weak values are said to be a ``new'' kind of physical quantity which ``can be interpreted as a form of conditioned average associated with an observable'' \cite{wv_review}.  Here we give an operational framework for this conditioned average and show that the anomalous values are a ubiquitous feature of generic statistical models not limited to quantum theory.  This places serious constraints on the explanatory power of weak values.

Why is such an operational framework important?  In 1964, Bell \cite{bell} described a statistical thought experiment which involved two stations which accepted binary inputs and produced binary events conditional on these inputs.  The setup contained \emph{absolutely no reference to physical theories} although was devised to teach us a lesson about physics.  This was necessary and crucial since the framework was designed to test theories themselves and thus was required to be model independent.  Bell showed two things with this remarkably simple set up: (1) an inequality on the value of a certain statistic calculated from the data was satisfied by \emph{every} local hidden variable model; and, (2) entangled quantum systems could violate this inequality.

The consequences of Bell work are innumerable.  By showing a separation between quantum and classical theory for an \emph{operational} task, Bell paved the way for considering quantum mechanical effects not as a nuisance, but as a resource.  Here we stand on the shoulders of Bell and propose an analogous operational framework for what is known as pre- and post-selected (PPS) measurement.  That is, we propose a model independent statistical experiment and give an inequality on a certain statistic derived from the data called the \emph{post-selected shift}, which can be thought of as an operational generalization of the \emph{weak value} from quantum theory.  The inequality is satisfied for any statistical model lacking a certain kind of correlation.  Since we brought up Bell, one might expect that the only a theory with correlations as rich as quantum mechanics might violate this inequality.  However, we show that classically correlated systems, even without disturbance, can violate the inequality.  We call the violation of the inequality an \emph{anomalous post-selected shift}.  Such an effect would, for example, be a necessary consequence of observing anomalous weak values in quantum mechanical systems.  If weak values are truly ``quantum'', this work proves it is not the anomalous shift of measurement pointers which signals it.   

\begin{figure*}\centering
  \includegraphics[width=0.95\textwidth]{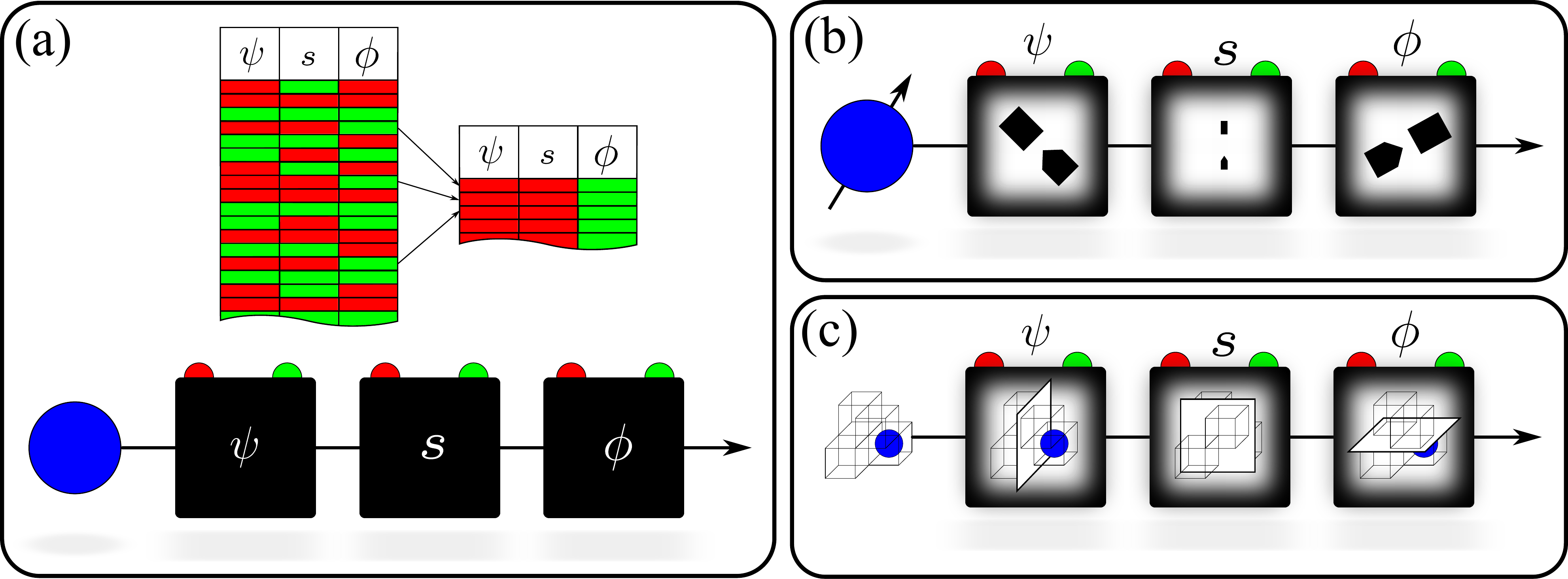}
  \caption{\label{all} (a) The general setup for three dichotomous events $\psi\to s\to \phi$.  In the text we label red as ``$+1$'' and green as ``$-1$'' (assume blue LEDs had not been invented yet).  Some exemplary data set is shown on the left and a pre- and post-selection (PPS) ensemble on the right, showing the \emph{anomalous post-selection shift} (APSS).  In this example, the unconditioned average of $s$ is zero while the post-selected average is maximal.  (b) a quantum mechanical \emph{weak measurement} experiment realizing APSS (weak values).  Inside the black boxes are Stern-Gerlach magnets, the first and last of which perform standard measurements while the middle performs a ``weak'' measurement of the spin-1/2 particle.  (c) a classical experiment realizing APSS.  From left to right, the measuring devices test left-vs-right, front-vs-back and top-vs-bottom to determine which of the 4 boxes the ball is in.}
\end{figure*}

The statistical model is that of three dichotomous events, $\psi$, $s$ and $\phi$.  We need not make any assumptions about the temporal ordering of these events but it is helpful to consider the sequence $\psi \to s\to \phi$. In doing so, we refer to the events as follows: $\psi$ as pre-selection, $s$ as the intermediate measurement, and $\phi$ as post-selection.  This situation is depicted in Fig.~\ref{all} (a).  The most important thing to remember is that there is a single object which encodes all the information we have about the situation: the joint probability
\begin{equation}
\Pr(\psi,s,\phi).
\end{equation}
This point cannot be emphasized enough and could potentially go a long way to resolving other statistically natured paradoxes in quantum theory so we will restate it: \emph{complete information about any statistical problem is contained in the joint probability distribution of the random variables}.
Every other probability distribution of interest can be derived from the rules of probability from this object.  A couple examples are:
\begin{align}
\Pr(\psi,\phi) &= \sum_s \Pr(\psi,s,\phi),\label{marginal}\\
\Pr(s|\psi,\phi) &= \frac{\Pr(\psi,s,\phi)}{\Pr(\psi,\phi)}\label{conditional}.
\end{align}

Without loss of generality, we take $\psi,s,\phi\in\{\pm 1\}$.  Considering the power series of any function of these variables, it is not difficult to see that any statistical model of these variables can be written
\begin{align}\label{gen}
\Pr(\psi,s,\phi) =& \frac{1}{8}\Big [c_0+c_1\psi+c_2 s+c_3\phi + c_4\psi s +c_5 \psi\phi +\nonumber \\
&\quad\,\, c_6 s\phi + c_7 \psi s\phi\Big].
\end{align}
Clearly, there must be constraints on the coefficients.  The normalization of the distribution, that is
\begin{equation}
\sum_{\psi,s,\phi} \Pr(\psi,s,\phi) = c_0,
\end{equation}
implies $c_0 = 1$ for all models.  By considering all possibilities of the $\pm 1$ values of the variables, it is easy to see that the positivity condition amounts to 8 inequalities which feature sums and differences of the coefficients, which we will not write out explicitly.  Because every model can be written in this form, a model is equivalently specified by its coefficient vector:
\begin{align}
\vec{c} = ( 1, c_1, c_2, c_3, c_4, c_5, c_6, c_7).
\end{align}

The experimentally accessible quantities we consider here are the pre-selected shift $\mathbb E_{s|\psi}[s]$ and the post-selected shift $\mathbb E_{s|\psi,\phi}[s]$. {Here we have used the notation that $\mathbb E_{x|y}[f(x)]=\sum_x \Pr(x|y) f(x)$ is the conditional expectation of $f(x)$ given the event $y$ }.  The term ``shift'' here refers to the case where $s$ represents, say, the coarse grained position of a physical meter.  Under naive classical intuition, the initial condition $\psi$ suffices to determine the full evolution of the state.  Thus we might expect
\begin{equation}\label{naive}
\left|\mathbb E_{s|\psi,\phi}[s]\right|\leq\left| \mathbb E_{s|\psi}[s] \right|.
\end{equation}
On the other hand, if it happens that
\begin{equation}\label{APSS!}
\left|\mathbb E_{s|\psi,\phi}[s]\right|>\left| \mathbb E_{s|\psi}[s] \right|,
\end{equation}
we call this an \emph{anomalous post-selected shift} (APSS).  In this work, we give a necessary condition for APSS and an explicit classical example---in some sense the simplest possible example---not requiring disturbance which achieves APSS.  First, however, we review the motivation for considering APSS from quantum mechanics, where the effect is referred to as anomalous \emph{weak values} \cite{Aharonov1988How}.

\section{Weak and strong quantum mechanical models}

\begin{figure}\centering
  \includegraphics[width=0.7\columnwidth]{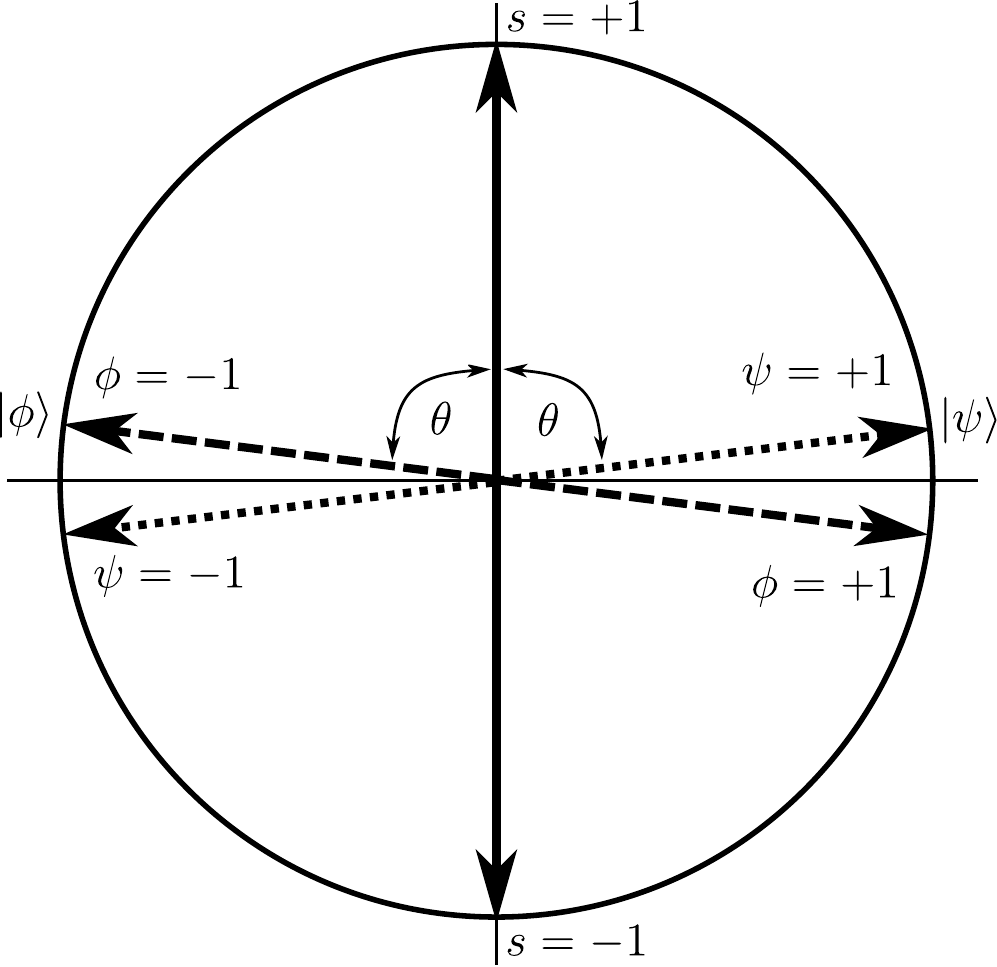}
  \caption{\label{fig} The Bloch sphere representation of the states involved in the strong and weak measurement protocol of quantum theory.}
\end{figure}

Consider the states 
\begin{equation}\label{states}
\ket{\psi} = \left(\begin{array}{c} \cos\theta/2\\\sin \theta/2\end{array}\right) \text{ and } \ket{\phi} = \left(\begin{array}{c} \phantom{-}\cos\theta/2\\-\sin \theta/2\end{array}\right),
\end{equation}
and their orthogonal complements{: $\ket{\psi_\perp}=(-\sin\theta/2, \cos\theta/2)^T$ and $\ket{\phi_\perp}=(\sin\theta/2, \cos\theta/2)^T $}.  Since there is only two possibilities for each, we label the states in Eq. (\ref{states}) as $\psi= +1$ and $\phi = -1$ with the orthogonal states given the opposite sign.  In the time between observing which of these states happen, we perform a measurement of the Pauli $Z$ operator, labeling the outcomes as $s=\pm 1$ for the corresponding eigenvalues.  Thus, we have three dichotomous random variables and the general model outline above applies.  These states and measurement operators are depicted in Fig.~\ref{fig}.  A schematic of an experiment realizing this setup is shown in Fig.~\ref{all} (b).

The Born rule dictates
\begin{equation}
\Pr(s|\psi) = |\langle s|\psi\rangle|^2 = \frac12 ( 1 + \cos\theta \psi s).
\end{equation}
Since the post-measurement state is projected into the eigenstates of $Z$, the final measurement probabilities are
\begin{equation}
\Pr(\phi|s,\psi) = \Pr(\phi|s) = |\langle \phi|s\rangle|^2 = \frac12 ( 1 - \cos\theta s\phi).
\end{equation}
Assuming the initial state is $\psi=\pm 1$ with equal probability, i.e. $\Pr(\psi)=1/2$, the joint distribution is then
\begin{align}
\Pr(\psi,s,\phi) &= \Pr(\phi|s,\psi)\Pr(s|\psi)\Pr(\psi),\\
& = \frac12 ( 1 - \cos\theta s \phi) \frac12 ( 1 + \cos\theta \psi s) \frac 12,\\
&= \frac18(1+\cos\theta \psi s -\cos^2\theta \psi \phi-\cos\theta s \phi).
\end{align}
In terms of the general model Eq.~\eqref{gen}, the strong measurement model is completely specified by coefficient vector
\begin{equation}\label{c_QS}
\vec{c}_{\rm QS} = (1,0,0,0,\cos\theta,-\cos^2\theta,-\cos\theta,0).
\end{equation}
The label QS will refer to this model---strong quantum measurement.

Now suppose we perform a \emph{weak} measurement of the Pauli $Z$ operator, which is described by the quantum operation
\begin{equation}
\mathcal E_s \rho =\frac{1}{2} \left [\rho  +s  \frac{\lambda }{2}\left (Z \rho + \rho Z\right ) \right ],
\end{equation}
where $0\leq \lambda\leq \cos\theta$ is the ``weakness'' parameter.  The conditional state is given by $\rho_s=\mathcal E_s \rho/ {\rm Tr}[{\mathcal E_s \rho}] $, and for our example $\Pr(s|\psi)={\rm Tr}[{\mathcal E_s \rho}]  = (1+ \lambda s\psi \cos\theta)/2$.
Using the conditional state above and $\Pr(s,\phi|\psi) = \Pr(\phi|s,\psi)\Pr(s|\psi)$, the joint distribution of $s$ and $\phi$ given $\psi$ is
\begin{align}
\Pr(s,\phi|\psi) & = \langle \phi|\mathcal E_s (\op \psi \psi )|\phi\rangle,\nonumber \\
& = \frac12 \langle \phi|  |\psi\rangle\!\langle \psi|  +\frac {s\lambda}{2}  ( |\psi\rangle\!\langle \psi| Z +  Z\op\psi\psi)  |\phi\rangle,\nonumber \\
&= \frac12\left(|\langle \phi|\psi\rangle|^2 + s\lambda \langle \phi|\psi\rangle \!\langle \psi|Z|\phi\rangle\right).
\end{align}
By considering all possibilities, it can be shown that the joint distribution is given by the coefficient vector
\begin{equation}\label{C_QW}
\vec{c}_{\rm QW} = (1,0,0,0,\lambda\cos\theta,-\cos2\theta,-\lambda\cos\theta,0).
\end{equation}
Here QW will refer to this weak quantum measurement model.  Notice the difference between the strong and weak model; the terms ($c_4$ and $c_6$) which contain $s$ are multiplied by $\lambda$ and $c_5$ changes in a subtle way. 

Now consider the pre- and post-selected averages in the weak measurement model:
\begin{align}
\mathbb E^{\rm QW}_{s|\psi}[s] &=\lambda\cos\theta\psi,\label{QW_pre}\\
\mathbb E^{\rm QW}_{s|\psi,\phi}[s] &= \frac{\lambda\cos\theta(\psi-\phi)}{1-\cos 2\theta \psi\phi}.
\end{align}
If $\phi$ and $\psi$ are different, we have
\begin{equation}
\mathbb E^{\rm QW}_{s|\psi,\phi=-\psi}[s]= \frac{\lambda\psi}{\cos\theta},
\end{equation}
and thus,
\begin{equation}\label{ineq_QW}
\left|\mathbb E^{\rm QW}_{s|\psi,\phi=-\psi}[s]\right|\geq\left| \mathbb E^{\rm QW}_{s|\psi}[s] \right|.
\end{equation}
This model possesses APSS.  In the quantum mechanical literature, this is the \emph{weak value} phenomenon.  Indeed,
\begin{equation}\label{QW_shift}
\mathbb E^{\rm QW}_{s|\psi,\phi=-\psi}[s]= \lambda z_w \psi,
\end{equation}
where, when $\phi=-\psi$,
\begin{equation}
z_w=\frac{ \langle \phi|Z|\psi\rangle}{ \langle \phi|\psi\rangle} = \frac{1}{\cos\theta}
\end{equation}
is called the \emph{weak value} \cite{Aharonov1988How}.  But, does Eq.~\eqref{ineq_QW} signify something uniquely quantum mechanical has happened?  Does it even have physical significance?  Is the weak value a physical quantity?  The proponents would have us believe the answer is, emphatically, yes.  

The reason is as follows.  One can view Eq.~\eqref{QW_pre} as the expectation value of the course-grained shift of a meter performing the measurement.  Notice that the value of the observable $Z$ is proportional to this shift:
\begin{equation}
\mathbb E^{\rm QW}_{s|\psi}[s]= \lambda \langle \psi|Z|\psi\rangle.
\end{equation}
But when we perform the pre- and post-selection, we have Eq.~\eqref{QW_shift}.  This appears as if the meter has shifted by an amount proportional to $z_w$ and, moreover, this shift can be larger than the expected shift without post-selection.  This had led some to suggest that $z_w$ is a new kind of physical property of quantum systems \cite{Aharonov1988How}.   

Be wary, however, that such ideas are guided by falsely applying physical intuition to logical inference which leads to the far worse problem of imposing physical meaning on logical deductions.  We hope by understanding the conditions under which APSS occurs, this deep misunderstanding of the purpose and role statistics plays in physical theories can be overcome.

As a first example, we need not look farther that the strong measurement model above.  In the strong measurement model, the expectation value of $s$ in the distributions of $s|\psi$ and $s|\psi,\phi$ are
\begin{align}
\mathbb E^{\rm QS}_{s|\psi}[s] &=\cos\theta\psi,\label{QS_shift}\\
\mathbb E^{\rm QS}_{s|\psi,\phi}[s] &= \frac{\cos\theta(\psi-\phi)}{1-\cos^2\theta \psi\phi}.
\end{align}
Notice that for $\psi = - \phi$ and any $\theta$, we have
\begin{equation}\label{ineq_QS}
\left|\mathbb E^{\rm QS}_{s|\psi,\phi=-\psi}[s]\right|\geq\left| \mathbb E^{\rm QS}_{s|\psi}[s] \right|.
\end{equation}
The strong measurement model possesses APSS as well.  Thus, the distinction between ``weak" and ``strong"  measurement is a red herring---APSS can occur with or without the notion of measurement strength.   Hence, in the remainder, we work with the fully general model \eqref{gen}.

\section{An inequality for conditionally independent models}
The defining feature of APSS is the reversal of the inequality in moving from Eq.~\eqref{naive} to Eq.~\eqref{APSS!}, as occurs in the quantum models above---see Eq.~\eqref{ineq_QW} and Eq.~\eqref{ineq_QS}.   Here we show that any model in which $s$ and $\phi$ are conditionally independent given $\psi$ cannot possess APSS.

\begin{thm}
In a general 3 variable dichotomous model \eqref{gen}, if $\Pr(s,\phi|\psi) = \Pr(s|\psi)\Pr(\phi|\psi)$, then
\begin{align}
c_6 &= c_2c_3+c_4c_5,\label{ind1}\\
c_7 &= c_2c_5+c_3c_4,\label{ind2}
\end{align}
and
\begin{equation}\label{thm}
\left|\mathbb E_{s|\psi,\phi}[s]\right|\leq\left| \mathbb E_{s|\psi}[s] \right|.
\end{equation}
\end{thm}
The proof is as follows.  Consider:
\begin{align}
\mathbb E_{s|\phi,\psi} \left[s\right]& = \sum_s s\Pr(s|\psi,\phi),\\
& =  \sum_s s\frac{\Pr(s,\phi |\psi)}{\Pr(\phi|\psi)},\\
& =  \sum_s s\Pr(s|\psi),\\
&=\mathbb E_{s|\psi} \left[s\right].
\end{align}
Thus, $\Pr(s,\phi|\psi) = \Pr(s|\psi)\Pr(\phi|\psi)$ implies Eq.~\eqref{thm}.  

Now, Eqs.~\eqref{ind1} and \eqref{ind2} are true for any model, but the equations become too unwieldy with all 8 coefficients.  Seeing the proof with $c_1=0$ (which halves the number of terms we have to consider in the expansions below) should make the general proof obvious.  We will proceed by comparing 
\begin{align}
&4\Pr(s,\phi|\psi) = \\
&c_0+c_2 s+c_3\phi + c_4\psi s +c_5 \psi\phi + c_6 s\phi + c_7 \psi s\phi\nonumber
\end{align}
to
\begin{align}
4\Pr(s|\psi)\Pr(\phi|\psi) =& (c_0+c_2 s+c_4\psi s)(c_0+c_3\phi +c_5 \psi\phi),\nonumber\\
 = &\Big[c_0^2+c_0c_2 s+ c_0c_3\phi + c_0c_4\psi s+\nonumber\\
 &\quad c_0c_5\psi \phi + ( c_2c_3+c_4c_5) s\phi +\nonumber \\
 &\quad (c_2c_5+c_3c_4)\psi s\phi \Big].
\end{align}
Matching coefficients reveals that the only two equations, which are not trivially satisfied, are Eqs.~\eqref{ind1} and \eqref{ind2}.  Note also that conditional independence is necessary but not sufficient.  In a correlated model, the post-selected shift is necessarily \emph{different} than the pre-selected shift---it just need not be \emph{larger}.  

\section{Classical violation without disturbance}
In Ref.~\cite{ferrie2} we devised a classical model specified by
\begin{equation}\label{c_class1}
\vec{c} = (1,0,0,\delta,\lambda,0,0,0),
\end{equation}
where $0<\delta<1-\lambda$ was interpreted in a physical model of coin toss as ``disturbance''.  The model gives  
\begin{equation}
\mathbb E_{s|\psi+1,\phi=-1}[s] = \frac{\lambda}{1-\delta} > \lambda =  \mathbb E_{s|\psi=+1}[s].
\end{equation}
A quick check also reveals Eq.~\eqref{ind2} is not satisfied.  Either of these facts imply the model is correlated, as required. However, let us be quick to note that physically interpreting the correlation as disturbance is not necessary.  

The simplest way to violate the inequality is to have all terms but one of $c_6$ or $c_7$ not equal to zero ($c_0$ necessarily equally one).  For example, take the model with only $c_7=1$:
\begin{equation}\label{c_class2}
\vec{c} = (1,0,0,0,0,0,0,1)
\end{equation}
or
\begin{equation}
\Pr(\psi,s,\phi) = \frac{1+ \psi s \phi}{8}.
\end{equation}
This model implies $\Pr(s|\psi,\phi) = (1+ \psi s \phi)/2$ and $\Pr(s|\psi) = 1/2$. Using these relations, notice that
\begin{equation}
1=\left|\mathbb E_{s|\psi,\phi}[s]\right|>\left| \mathbb E_{s|\psi}[s] \right| = 0,
\end{equation}
a maximally anomalous shift!  A physical realization of this model is shown in Fig.~\ref{all} (c).  A ball is placed in one of six boxes which are arranged in a tetrahedral pattern.  The measuring devices can ask the binary questions shown in the figure (left-vs-right, front-vs-back and top-vs-bottom).  So, for example, when we first find ``right'', we still do not know whether the ball is in the front or back.  But if the final measurement finds ``bottom'', necessarily the intermediate measurement gave ``back''.  Indeed,
\begin{equation}
\Pr(s|\psi=+1,\phi=-1) = \frac{1-s}{2}.
\end{equation}
That is, the pre-selected average is zero while the post-selected average is $-1$.  This is not surprising---more data ought to change ones knowledge of unknown quantities.  Moreover, this (rather large) difference does not signify a physical change in the system itself, only our knowledge of it.

\section{Discussion}
The last example shows beyond any doubt that the APSS effect is revealing the correlations present in the model.  The example is unequivocally classical by any measure.  Observing APSS therefore does not imply anything more exotic than classical statistics.  In particular, APSS does not bear any relation to physical notions, such as amplitudes \cite{kastner}, invasiveness \cite{WillJor08, DreJor12}, disturbance \cite{asger}, contextuality \cite{pusey} or interference \cite{dressel}.  These properties are all consequences---interesting ones, no doubt---of the structure of quantum theory, some of which are present in various classical theories.  But we have shown here that if that task is to observe an anomalous post-selection shift, the structure of quantum theory is not required.

What is missing for most of the proposed physical meanings for weak values is an understanding of how they arise as a resource for some operational task. Though, another example of an operational inequality (derived in the context of so-called Leggett-Garg inequalities) has been given that holds for any hidden variable model without disturbance \cite{LGI}.  A similar result would hold here on a hidden variable view.  In this context, the role of the hidden variables is to provide a ``common cause'' for the correlations between $s$ and $\phi$, given $\psi$.  Such a cause would render $s$ and $\phi$ conditionally independent given both $\psi$ and the posited hidden variable.  Such a demand would require that the consequence of our theorem [Eq.~\eqref{thm}] always holds.  In other words, within a hidden variable model, a change in the hidden variable in moving from the intermediate to post-selected measurement---disturbance---would be necessary to observe APSS.

Measurement is both a fundamental conceptual and practical tool in all areas of physics, and science more broadly.  One of the core consequences of quantum mechanics was in forcing us to rethink the meaning of measurement.  But some of the cherished niceties of classical physics that we had to give up were never even present in other branches of science.  Many branches of science do not even have the kinds of fixed and rigid models we take for granted in physics, and hence the notions of observables or physical quantities are irrelevant.  As a consequence, the word ``classical", when used in fundamental studies of quantum theory, must be taken with a grain of salt---is it really the most general classical argument to compare to?  Or, a more revealing question, can such claims be verified by an experiment provably separating theories?

Here we have given a simple probabilistic model which reveals anomalous shifts upon post-selection not requiring disturbance.  A theme in quantum foundations and information research is to develop classical models for aspects of quantum theory (e.g, \cite{speks}).  In this way, we hope to whittle down all the features of quantum theory until we are left with the kernel that is ``truly quantum''.  Can the correlations which cause anomalous weak values be considered a uniquely quantum feature?  The above argument casts doubt on this.

\begin{acknowledgements}
\textbf{Acknowledgements}---The authors thank Justin Dressel, Asger Ipsen, and {Ruth Kastner for helpful comments and discussions}.  This work was supported in part by National Science Foundation Grant Nos. PHY-1212445 and by the Canadian Government through the NSERC PDF program. JC was also supported in part by CERC, NSERC, and FXQI.
\end{acknowledgements}


\begin{thebibliography}{10}


\bibitem{tukey}
John W. Tukey, \emph{Sunset Salvo}, \href{http://www.jstor.org/stable/2683137}{The American Statistician {\bf 40} 72 (1986)}

\bibitem{wv_review}
Justin Dressel, Mehul Malik, Filippo M. Miatto, Andrew N. Jordan and Robert W. Boyd, {\em Colloquium: Understanding quantum weak values: Basics and applications}, \href{http://dx.doi.org/10.1103/RevModPhys.86.307}{Reviews of Modern Physics {\bf 86}, 307 (2014)}.

\bibitem{bell} 
J.S.~Bell, {\em On the Einstein Podolsky Rosen Paradox}, Physics {\bf 1} 195 (1964).

\bibitem{Aharonov1988How}
Yakir Aharonov, David Z. Albert and Lev Vaidman, \emph{How the result of a measurement of a component of the spin of a spin-1/2 particle can turn out to be 100},
  \href{http://dx.doi.org/10.1103/physrevlett.60.1351}{{Physical Review Letters} \textbf{{60}},
  {1351} ({1988})}.


\bibitem{ferrie2}
Christopher Ferrie and Joshua Combes, \emph{How the Result of a Single Coin Toss Can Turn Out to be 100 Heads}, \href{http://dx.doi.org/10.1103/PhysRevLett.113.120404}{Physical Review Letter {\bf 113}, 120404 (2014)}.

\bibitem{kastner}
R.E. Kastner, {\em Weak values and consistent histories in quantum theory}, \href{http://dx.doi.org/10.1016/j.shpsb.2003.02.001}{Studies in History and Philosophy of Science Part B: Studies in History and Philosophy of Modern Physics {\bf 35}, 57 (2004)}.

\bibitem{WillJor08}
Nathan S. Williams and Andrew N. Jordan, \emph{Weak Values and the Leggett-Garg Inequality in Solid-State Qubits}, \href{http://dx.doi.org/10.1103/PhysRevLett.100.026804}{Phys. Rev. Lett. {\bf 100}, 026804 (2008)}.

\bibitem{DreJor12}
J. Dressel and A. N. Jordan, \emph{Contextual-value approach to the generalized measurement of observables}, \href{http://dx.doi.org/10.1103/PhysRevA.85.022123}{Physical Review A {\bf 85}, 022123 (2012)}.


\bibitem{asger}
Asger C. Ipsen, \emph{Disturbance in weak measurements and the difference between quantum and classical weak values}, \href{http://arxiv.org/abs/1409.3538}{arXiv:1409.3538 (2014)}.


\bibitem{pusey}
Matthew F. Pusey, \emph{Anomalous weak values are proofs of contextuality}, \href{http://arxiv.org/abs/1409.1535}{arXiv:1409.1535}.

\bibitem{dressel}
Justin Dressel, \emph{Weak Values are Interference Phenomena}, \href{http://arxiv.org/abs/1410.0943}{arXiv:1410.0943}




\bibitem{LGI}
J. Dressel, C. J. Broadbent, J. C. Howell and A. N. Jordan, {\em Experimental Violation of Two-Party Leggett-Garg Inequalities with Semiweak Measurements}, \href{http://dx.doi.org/10.1103/PhysRevLett.106.040402}{Physical Review Letters {\bf 106}, 040402 (2011)}.


\bibitem{speks} Robert W. Spekkens, \emph{Evidence for the epistemic view of quantum states: A toy theory}, \href{http://dx.doi.org/10.1103/PhysRevA.75.032110}{Physical Review A {\bf 75}, 032110 (2007)}.

\end{thebibliography}
\end{document}